\def\CF{C_{\!_F}}
\def\CA{C_{\!_A}}
\def\TF{T_{\!_F}}
\def\BETAzero{\beta_{\!_0}}
\def\BETAone{\beta_{\!_1}}
\def\Euler{\gamma_{\!_E}}

\def\Rbar{\,\,\overline{\!\!\hskip1pt{R}\!\hskip1.5pt}}
\def\vbar{\overline{v\!\hskip1.5pt}}
\def\mbar{\,\overline{\! m}\hskip1pt}
\def\qqq{{q}}
\def\qbar{\bar{q}}

\def\fbar{\,\,\overline{\!\!f}}
\def\cbar{\,\overline{\!c\hskip1pt}}
\def\bbar{\bar{b}}
\def\tbar{\bar{t}}

\documentstyle[12pt,eqsecnum,preprint,prd,aps,epsf]{revtex}
\begin{document}
\draft
\preprint{FTUAM 95--2}
\date{February, 1995}
\title{Production of Heavy Quarks Close to
Threshold\footnote{Research partially supported by CICYT,
Spain.}}
\author{K.~Adel\footnote{Electronic address: {\tt
kassa@nantes.ft.uam.es.}} and F.~J.~Yndur\'ain\\
Departamento de F\'{\i}sica Te\'orica, C-XI,\\ Universidad
Aut\'onoma de Madrid, Canto Blanco,\\ E--28049 Madrid,
SPAIN}
\maketitle
\begin{abstract}
We calculate production by vector and axial currents of heavy quark
pairs ($c{\cbar}$, $b{\bbar}$, $t{\tbar}$) close to threshold.  We
take into account strong interaction contributions (including
radiative corrections and leading nonperturbative effects) by using
the Fermi-Watson final state interaction theorem. We use the results
obtained to compare with experiment for open production of $c{\cbar}$,
$b{\bbar}$ near threshold, and to give a reliable estimate of the
so-called ``threshold effects'' contribution to vector and
axial correlators, for $t{\tbar}$, $i.e.$, the contribution of regions
close to $\,4\,m_t^2\,$ to $\Pi(t)$, for small values of $t$
($0<t\;\lower2pt\hbox{$\lesssim$}\; M_Z^2$ ).
\end{abstract}
\pacs{14.40.Gx, 12.38.Bx, 12.38.Lg, 13.20.Gd}
\section{Introduction}
\label{sec:1}
In this paper we consider the production of a pair
$q{\qbar}$ of heavy quarks close to threshold; that is to
say, for small values of $|v|$ where
\begin{equation}
\eqnum {1.1}
v \equiv \sqrt{1-4m^2/s} \ .
\end{equation}
Here $m$ is the quark mass, and $s^{1/2}$ the center of mass
energy of the $q{\qbar}$ pair.  Above threshold $v$
coincides with the velocity of either quark; but we also
study the production of $q{\qbar}$ bound states.

We take the quarks to be heavy, so that we may be able to
apply a perturbative QCD analysis to them.  Thus we study
production of $c{\cbar}$, $b{\bbar}$ and $t{\tbar}$. (For
the last case, and in this first paper, we will neglect the
effects of $t$ decay). With a view to applications to
production by $e^+e^-$ collisions, we take the $q{\qbar}$ to be
produced by either a vector $V$ or axial $A$ current\footnote{Our
results may be extended with little effort to scalar or pseudoscalar
correlators.}:
\begin{eqnarray}
\nonumber
V(x) &=& {\qbar}(x) \; \gamma_{\mu} \; q(x) \\
\eqnum {1.2}
A(x) &=& {\qbar}(x) \; \gamma_{\mu} \gamma_5\;  q(x).
\end{eqnarray}
Sum over omitted color indices is understood.  We will then
study the correlators,
\begin{eqnarray}
\nonumber
\Pi_{\mu\nu}^V (p) &=& i \int d^4x \; e^{ip\cdot x}
	\langle vac|\ T\, V_{\mu}(x) V_{\nu}(0) \ |vac \rangle \\
\eqnum {1.3}
 		   &=& (-p^2 g_{\mu \nu} + p_{\mu} p_{\nu})\,\Pi_V(p^2)\ ,\\
\nonumber
\Pi_{\mu\nu}^A (p) &=& i \int d^4x \; e^{ip\cdot x}
	\langle vac|\ T\, A_{\mu}(x) A_{\nu}(0) \ |vac \rangle \\
\eqnum {1.4}
		   &=& (-p^2 g_{\mu \nu} + p_{\mu} p_{\nu})\,\Pi_A(p^2)
			+ p_{\mu} p_{\nu}\,\Pi_P(p^2)\ ,
\end{eqnarray}
where $|vac\rangle $ denotes the physical vacuum.  As is
known, the production cross section for $e^+e^- \rightarrow
q{\qbar}$ may be straightforwardly written in terms of the
imaginary parts of the correlation functions
\begin{equation}
 {\rm{Im}}\,\Pi(p^2)\ \ , \hskip1truecm \Pi = \Pi_{V,\,A,\,P}
\ \ ; \eqnum {1.5}
\end{equation}
only $\Pi_V$, $\Pi_A$ give sizeable contributions, and they
are the quantities we will study here.

To lowest order in perturbation theory ($i.e.$, the parton
model) we may neglect the interactions of $q{\qbar}$.  The
$\Pi$ are then obtained with a simple one-loop evaluation
and one has
\begin{eqnarray}
\eqnum{1.6.a}
&& {\rm{Im}}\,\Pi_V^{(0)}(s) = { N_c \over 12\,\pi}
\; {v(3-v^2) \over 2} \; \theta(s-4\,m^2) \ \ , \\
\eqnum{1.6.b}
&& {\rm{Im}}\,\Pi_A^{(0)}(s) = { N_c \over 12\,\pi}
\; v^3 \; \theta(s-4\,m^2) \ \ .
\end{eqnarray}
Here $N_c= \# \ {\rm{colors}} \ =3$, and the subscript zero
in the $\Pi$ indicates that the strong interactions are
neglected.  Of course strong interactions are most important
near threshold: its incorporation is precisely the subject
of the present paper.

This article is organized as follows. In section 2 we give
the expression for the contribution of the $q{\qbar}$ bound
states to the ${\rm{Im}}\,\Pi$.  This, we hope, will serve
to clear some of the misunderstandings found in the standard
literature. We also deal, in this section, with the
${\rm{Im}}\,\Pi(s)$ above threshold, but in the
nonrelativistic regime ($v^2 \ll 1$), explaining the use of
the final state interaction theorem to incorporate strong
interactions, which again should clarify some of the
existing fog.

In section 3, we use the known evaluations of wave function
for bound states, and the one obtained here in the
continuum, to give explicit formulas for
${\rm{Im}}\,\Pi(p^2)$ below threshold and above.  In this
last case radiative and nonperturbative contributions are
included for the first time. The article is concluded in
section 4, where we discuss in detail the important case of
${\rm{Im}}\,\Pi_V(p^2)$ above threshold and the contribution
of this region to the evaluation of $\Pi_V(q^2)$ for $q^2 \ll
m^2$. An Appendix is also provided for the technical details
of the calculations.

\section{The imaginary part of $\Pi$ around threshold}
\label{sec:2}
\subsection{Bound state contributions to ${\rm{Im}}\; \Pi$.}
\label{subsec:2.1}
We will carry over the detailed calculations for ${\rm
{Im}}\; \Pi_V$; then we will indicate the corresponding
results for ${\rm {Im}}\; \Pi_A$.  If we have a bound state
of $q{\qbar}$ with momentum $k$, $k^2=M^2$, and third
component of spin $\lambda$, normalized to
\begin{displaymath}
\langle k,\lambda \;|\; k', \lambda' \rangle = 2\,k_0\,\delta_{\lambda
\lambda'} \,
\delta ( \vec{k} - \vec{k}' )\ ,
\end{displaymath}
then its contribution to ${\rm {Im}}\; \Pi_{\mu \nu}^V(p)$ is
\begin{eqnarray}
\nonumber
 {\rm {Im}}\; \Pi_{\mu \nu}^{V;\;pole} (p) &=& {1 \over 2} \int d^4x \;
e^{i\,p\cdot x}
\sum_{\lambda} \int {d^3k \over 2 \,k_0} \;
\langle vac|\, J_{\mu}(x) \,|k,\lambda\rangle \; \langle k,\lambda|
\, J_{\nu}(0)\, |vac\rangle \\
\eqnum {2.1}
&=& {1\over 2} (2\pi)^4 \delta (p^2-M^2) \sum_{\lambda} \;
	V_{\mu}(p,\lambda) \; V_{\nu}^*(p,\lambda) \ \ .
\end{eqnarray}
In the above equation,
\begin{equation}
V_{\mu}(p,\lambda) \equiv
\langle vac| \;{\qbar} (0) \gamma_{\mu} {q} (0) \;|p,\lambda\rangle
\eqnum {2.2}
\end{equation}
is the ($p$-space) wave function of the bound state by
{\em{definition}}.  The connection with the $x$-space wave
function may be carried over immediately in the
{\em{nonrelativistic\/}} limit.  In the c.m. referencial one
thus finds, for $e.g.$ $\lambda=+1$, and with $k=p_1+p_2$,
\begin{eqnarray}
\nonumber
 \langle vac| \; {\qbar} (0) & & \gamma_{\mu} {q} (0) \;
|k,\lambda=1\rangle \\
\eqnum {2.3}
& &=\  { \sqrt{N_c} \over (2\,\pi)^{3/2} \; }\;
\sqrt{ { k_0 \over  2 \, p_{10} \, p_{20} } } \;
{\vbar}(p_1,1/2) \; \gamma_{\mu} \; u(p_2,1/2) \; \Psi(0) \ .
\end{eqnarray}
Here $\Psi(0)$ is the $x$-space wave function evaluated at
$\vec{r}=0$: thus, only states with $\ell=0$ contribute.

It is perhaps not idle to note that Eqs. (2.1, 2.3) are
exact, the last if taken as a definition of $\Psi(0)$.
Substituing one into the other, we finally obtain
\begin{equation}
{\rm {Im} } \, \Pi_V^{pole} (p^2) = { N_c \over M} \; \delta(p^2-M^2)
\; \left| R_0(0) \right|^2 \; , \eqnum {2.4.a}
\end{equation}
and $R_0 = (4\,\pi)^{1/2} \, \Psi_{\ell=0}$ is the radial wave function.

For the axial correlator the evaluation is slightly more
complicated because the orbital angular momentum of the
bound states is $\ell=1$. One finds,
\begin{eqnarray}
\nonumber
&& {\rm {Im} } \; \Pi_A^{\,pole} (p^2) = { 3\,N_c \over 2\,m^2\,M} \;
\delta(p^2-M^2) \; \left| R\,^{\,'}_1(0) \right|^2 \; ,  \\
\eqnum {2.4.b}
&& R\,^{\,'}_1(r) = \partial R_1 (r) \, / \, \partial r \ \  ;
\end{eqnarray}
$R_1$ is the radial part of the $\ell=1$ wave function. In
the nonrelativistic limit the $R_{\ell}$ are normalized to
\begin{displaymath}
\int_0^{\infty} dr \; r^2 \left|\, R_{\ell}(r) \; \right|^2 = 1 \ .
\end{displaymath}
%
\subsection{$\ {\rm{Im}}\,\Pi$ above threshold.}
\label{subsec:2.2}
We will give a detailed discussion for the case where we
have the vector correlator. Moreover, we will consider only
the production via a virtual photon, neglecting the
contribution of the $Z$.  (This only for ease of discussion;
the results will be valid quite generally.)  In these
circumstances the quantity ${\rm{Im}}\;\Pi_{\mu\,\nu}^V$ may
be considered, for $p^2 > 4\,m^2$, to be proportional to the
square of the production amplitude $\gamma^* \rightarrow
q{\qbar}\; $:
\begin{displaymath}
{\rm{Im}}\;\Pi_{\mu\nu} \sim \big| \;
\langle q{\qbar} | \; S \; |\gamma^* \rangle \; \big|^2 \ \ \ .
\end{displaymath}
We will consider that the interactions involved in this
process are two: the electromagnetic
\begin{displaymath}
 H_{I\,{\rm{em}}} \ =\  e\, Q_f
\int d\,^3\!x \; {\qbar}(x) \,\gamma_{\mu} \, q(x) \, A^{\mu}(x)
\ \ \ ,
\end{displaymath}
and the QCD interaction described by a Hamiltonian
$H_{IQCD}$ that will be specified later. The final state
interaction theorem then asserts that the amplitude $\langle
q{\qbar}|\,S\,|\gamma^*\rangle $ may be evaluated, to first
order in $H_{I\,{\rm{em}}}$, but {\em{to all orders\/}} in
$H_{IQCD}$, by means of the expression
\begin{equation}
\langle q{\qbar}|\,S\,|\gamma^*\rangle  \ =\ i \langle \Psi| \;H_{I\,{\rm{em}}}
\;|
\gamma^*\rangle  \ \ , \eqnum {2.5}
\end{equation}
where $|\Psi\rangle $ is an {\em{exact\/}} solution of the
Lippmann-Schwinger equation for $q{\qbar}$ states subject to
the strong interaction:
\begin{equation}
|\Psi\rangle  = |\Psi^{(0)}\rangle  + { 1 \over E - H_0 }\; H_{IQCD} \
|\Psi\rangle
\ \ . \eqnum {2.6}
\end{equation}
(In our case, and because only one wave contributes, we need
not specify the boundary conditions in Eq. (2.6)).

There is now a complication, as compared to the bound state
case: $|\Psi\rangle $ may now contain, besides $q{\qbar}$,
$q{\qbar}+n$ gluons.  To the order we will be working and in
the nonrelativistic approximation , the states $|q{\qbar}+n\
{\rm{gluons}}\rangle $ may however be neglected.  The reason
is that the amplitude for radiation of a gluon by
{\em{heavy}} quarks is proportional to the velocities $\ v_q
+ v_{{\qbar}}\ $, so that the contribution of these
processes to ${\rm{Im}}\;\Pi_{\mu\nu}$ will be of order
$(v_q + v_{{\qbar}})^2 \sim v^2$, $i.e$, of the order of the
relativistic corrections, which we are neglecting.
Therefore, in the {\em{nonrelativistic}} limit the state
$|\Psi\rangle $ may be considered to consist only of
$q{\qbar}$, and can thus be represented by a wave function.
We then obtain
\begin{equation}
{\rm{Im}}\;\Pi_V(s) = \big| \, R_{k0}(0) \, \big|^2 \;
{\rm{Im}}\;\Pi_V^{(0)}(s) \ \ , \eqnum {2.7.a}
\end{equation}
and also
\begin{equation}
  {\rm{Im}}\;\Pi_A(s) = \big| \, 3\,R\,^{\,'}_{k1}(0)/m\,v \, \big|^2 \;
{\rm{Im}}\;\Pi_A^{(0)}(s) \ \ . \eqnum {2.7.b}
\end{equation}
Here $R_{k\ell}$ is the radial part of the continuum wave
function with $k=m\,v$, and normalized to
\begin{equation}
\int_0^{\infty} dr \; r^2 \, R^*_{k\ell}(r) \, R_{k'\ell}(r) =
{2\,\pi \over k^2} \;\delta(k-k')  \ \ .\eqnum {2.8}
\end{equation}
The ${\rm{Im}}\;\Pi^{(0)}$ are as given in Eq. (1.6). Again
we would like to emphasize that Eqs. (2.7) are exact (in the
nonrelativistic limit); approximations enter when evaluating
the $R_{k\ell}$, which will be the subject of the next
section.

\section{The $\qqq\qbar$ wave function close to threshold}
\label{sec:3}
\subsection{The QCD interaction for heavy quarks at short distances.}
\label{subsec:3.1}
It has been known for a long time that the {\em{short
distance}} interactions of a pair of heavy quarks may be
described by perturbation theory; the {\em{leading}}
nonperturbative corrections are then implemented taking into
account the nonzero values of quark and gluon condensates in
the physical vacuum:
\begin{eqnarray*}
&& \langle q{{\qbar}}\rangle  \ \equiv \
\langle vac| \; :{{\qbar}}(0) \,q(0):\, |vac\rangle \ \ ,\\
&& \langle \alpha_{\hskip-1.pt s}\,G^2\rangle  \ \equiv\  \alpha_{\hskip-1.pt
s}  \langle vac| \ :G_{\mu\nu}(0)\,
G^{\mu\nu}(0): \ |vac\rangle  \ \ .
\end{eqnarray*}
The details may be found in the classical SVZ papers$^{[1]}$
where the correlators $\Pi_{\mu\nu}$ are directly studied
using the operator product expansion techniques; or the work
of Leutwyler$^{[2]}$ and Voloshin$^{[3]}$ where the Green's
function method is employed to study the $\;q{\qbar}\;$
bound states.  In particular, these last authors explicitly
prove that no potential may describe the {\em{short
distance}} nonperturbative corrections to the $\;q{\qbar}\;$
spectrum and wave function; instead, one has to use an
effective interaction which in the nonrelativistic limit is
given by
\begin{equation}
H_{INP} = - g \, \vec{r} \, \vec{\cal{E}}_a \,t^a \ \ , \eqnum {3.1}
\end{equation}
with ${\cal{E}}$ the chromoelectric field. One then takes,
\begin{displaymath}
\langle vac|\; \vec{\cal{E}} \; |vac\rangle  = 0 \ \ , \hskip0.6truecm
\langle vac|\; \vec{\cal{E}}\,^2 \; |vac\rangle  \ \sim \ \langle
\alpha_{\hskip-1.pt s} \,G^2 \rangle \ .
\end{displaymath}
(In our case the contribution of the quark condensate is
negligible).  From a practical point of view it has been
made apparent in the detailed evaluations of Ref. [4] that a
calculation of $\,q{\qbar}\,$ states, based on perturbation
theory, and supplemented by the leading nonperturbative
corrections, as given by Eq. (3.1), yields an excellent,
essentially parameter--free description of the bound states
of $\;c{\cbar}\;$ with $n=1$ ($n$ being the principal
quantum number) and of $\;b{\bbar}\;$ states with $n=1,2 \ ;
\ \ell=0,1$. As was already known from the work of Refs.
[2,3], the analysis breaks down for higher excited states
where nonperturbative contributions get out of hand and
calculation from first principles becomes impossible. This
occurs for $n>1$ in $\;c{\cbar}\;$, and $n>2$ for
$\;b{\bbar}\;$. For $\;t{\tbar}\;$ the distances involved
are so short that a rigorous calculation becomes possible up
to $n\sim 5$.

Besides the nonperturbative contributions described by Eq.
(3.1) we require also the interaction deduced using
perturbation theory.  In the nonrelativistic regime, and
including one-loop corrections\footnote{The radiative
corrections to the $q{\qbar}$ potential have been obtained
by a number of authors; cf. Ref. [5] and Ref. [4] where they
are completed and summarized.}, we have the Hamiltonian
\begin{eqnarray}
\nonumber
&& H = H_{eff}^{(0)} + H_1 \ \ ,\\
\eqnum {3.2}
&& H_{eff}^{(0)} = \; -\; {1\over m} \Delta \;-\;
{ \CF \widetilde{\alpha}_{\hskip-1.3pt s} (\mu^2) \over r} \ \ ,\\
\nonumber
&& H_1 = \;-\; {\CF \BETAzero \alpha_{\hskip-1.pt s}^2 \over 2 \pi}\;
{\ln r\mu \over r}\ \ ;
\end{eqnarray}
here $\widetilde{\alpha}_{\hskip-1.3pt s}$ includes part of the radiative
corrections,
\begin{eqnarray*}
&& \widetilde{\alpha}_{\hskip-1.3pt s}(\mu^2) = \alpha_{\hskip-1.pt s}(\mu^2)
\;
\left[ 1 + { a_1 + \Euler \BETAzero/2 \over \pi } \;
	\alpha_{\hskip-1.pt s}(\mu^2) \right] \ \ ;\\
&& \BETAzero = {11 \,\CA - 4 \,\TF \,n_f \over 3 } \ \ ,
\ \ a_1= { 31 \,\CA - 20 \,\TF \,n_f \over 36} \ \ ; \ \ \CF = {4 \over 3}, \
\CA=3, \ \TF={1 \over 2} \ \ ,
\end{eqnarray*}
and $n_f$ is the number of quark flavors with masses much
smaller than $m$.  The reason why we include part of the
radiative corrections in $H_{eff}^{(0)}$ is that this
Hamiltonian, being Coulombic, may (and will) be solved
exactly, whereas $H_1$ and $H_{INP}$ have to be incorporated
in perturbation theory.

A last point about Eq. (3.2) is the meaning of the parameter
$m$ there.  As follows from the analysis of Ref. [4], this
$m$ has to be interpreted as the {\em{pole}} mass. That is
to say, if $\displaystyle  S(\not\!p)$ is the quark propagator, in
perturbation theory, then $m$ is such that $\displaystyle  S^{-1}
\left( \not\!p = m \right) = 0 $. This $m$ may be related to
the $\overline{\rm{MS}}$ mass, ${{\mbar}} ( {\mbar}^2 )$ through
the formula$^{[6]}$
\begin{equation}
m = {{\mbar}} ( {\mbar}^2 ) \left\{
1 + { \CF \alpha_{\hskip-1.pt s}(m^2) \over \pi }
 + ( K - 2 \,\CF ) \left[ {\alpha_{\hskip-1.pt s}(m^2) \over \pi } \right]^2 +
\dots \right\}
\ .\eqnum {3.3}
\end{equation}
Here $K \sim 13.5$ (an exact formula for $K$ may be found in
Ref. [6]).  As for the {\em{numerical}} values of the
masses, the analysis of Ref. [4] gives
\begin{equation}
m_c = 1570 \pm 60 \ MeV \ \ \ ; \ \ \
m_b = 4906 \pm 85 \ MeV\ \ ,\eqnum {3.4}
\end{equation}
which corresponds to the $\overline{\rm{MS}}$ masses
\begin{displaymath}
{\mbar}_c({\mbar}_c^2) = 1306 \pm 40 \ MeV\ \ \ ,\ \ \
{\mbar}_b({\mbar}_b^2) = 4397 \pm 40 \ MeV\ \ \ ,
\end{displaymath}
of course compatible with (but more precise than) the values
obtained with the SVZ method$^{[1,7]}$
\begin{displaymath}
{\mbar}_c({\mbar}_c^2) = 1270 \pm 50 \ MeV\ \ \ ,\ \ \
{\mbar}_b({\mbar}_b^2) = 4250 \pm 100 \ MeV \ \ .
\end{displaymath}
We will thus take Eq. (3.4) as our input. For the top quark
we elect to choose
\begin{equation}
m_t = 165 \pm 15 \ GeV \ \ . \eqnum {3.5}
\end{equation}
%

\subsection{The bound state wave functions.}
\label{subsec:3.2.}
The energy levels calculated with Eqs. (3.2), (3.1)
are$^{[4]}$
\begin{eqnarray}
\nonumber
E_{n\ell} = \ 2m \;\Bigg\{\;  1 - &&
	{\CF^2 \,\widetilde{\alpha}_{\hskip-1.3pt s}(\mu^2)^2 \over 8\,n^2 }
-  { \CF \BETAzero \alpha_{\hskip-1.pt s}^2 \widetilde{\alpha}_{\hskip-1.3pt s}
\over 8 \,\pi\,n^2}
	\left[ \ln { \mu\,n \over m \CF \widetilde{\alpha}_{\hskip-1.3pt s} } +
\psi(n+\ell+1)
	\right] \\
\eqnum {3.6}
&& + \; { \pi \,\epsilon_{n\ell} \, n^6 \langle \alpha_{\hskip-1.pt s}
G^2\rangle  \over
	2 \left(m \CF \widetilde{\alpha}_{\hskip-1.3pt s} \right)^4 }
		\;\Bigg\} \ .
\end{eqnarray}
Here the $\epsilon$ are numbers of order unity:
\begin{displaymath}
\;\epsilon_{10} = {624\over 425}\ ;\ \ \ \epsilon_{20} = {1051\over 663}\ ;
\ \ \ \epsilon_{30} = {769456\over 463239} \  ; \ \ \
\epsilon_{21} = {9929\over 9945} \ ; \dots
\end{displaymath}
other $\epsilon_{n\ell}$ may be found in Ref. [4], and an
analytic expression in Ref. [2]. For the wave functions, the
details are given in Ref. [3] and, especially, in the second
paper of Ref. [4]. One has
\begin{eqnarray}
\nonumber
&& {\Rbar}_{n\ell}(r) = { 2 \over n^2 \, a(n,\ell)^{3/2} } \;
\sqrt{ { (n-\ell-1)! \over (n+\ell)! }} \, \rho_{n\ell}^{\ell} \;
\displaystyle  e^{-\rho_{\!n\ell}/2} \; L_{n-\ell-1}^{2\ell+1} ( \rho_{n\ell} )
\ , \\
\eqnum {3.7}
&& \rho_{n\ell} = 2\,r/a(n,\ell) \ \ ;\\
\nonumber
&& a(n,\ell) = { 2 \over m \CF \widetilde{\alpha}_{\hskip-1.3pt s} (\mu^2) }
\left\{
1 - { \ln (n\mu / m \CF \widetilde{\alpha}_{\hskip-1.3pt s}) + \psi(n+\ell+1) -
1 \over 2 \pi }\;
\BETAzero \alpha_{\hskip-1.pt s} \right\} \  .
\end{eqnarray}
This includes the radiative corrections, as obtained using Eq. (3.2).
The full wave function at the origin, including
nonperturbative contributions (Eq. (3.1)) is
\begin{equation}
R_{n\ell}(0) = \bigg(1+\delta_{NP}(n,\ell) \bigg)\; {\Rbar}_{n\ell}(0)
\eqnum {3.8}
\end{equation}
and, for the first $n,\;\ell$,
\begin{eqnarray*}
&& \delta_{NP}(1,0) =
\left\{ {2\,968 \over 425} + {968\,576 \over 541\,875} \right\}\;
{\pi \langle \alpha_{\hskip-1.pt s} G^2\rangle  \over m^4 (\CF
\widetilde{\alpha}_{\hskip-1.3pt s})^6 } \ ,\\
&& \delta_{NP}(2,0) =
 \left\{ {3\,828\,736 \over 1\,989} + {753\,025\,024 \over 1\,318\,707}
\right\}\;
{\pi \langle \alpha_{\hskip-1.pt s} G^2\rangle  \over m^4 (\CF
\widetilde{\alpha}_{\hskip-1.3pt s})^6 } \ ,\\
&& \delta_{NP}(2,1) =
 \left\{ {3\,299\,840 \over 1\,989} + {33\,026\,904\,064 \over 98\,903\,025}
\right\}\;
{\pi \langle \alpha_{\hskip-1.pt s} G^2\rangle  \over m^4 (\CF
\widetilde{\alpha}_{\hskip-1.3pt s})^6 } \ .
\end{eqnarray*}
Higher $\delta_{NP}(n,\ell)$ may be found in the second
paper of Ref. [4].

\subsection{The $\;q{\qbar}\;$ wave function in the continuum.}
\label{subsec:3.3}
The calculation of the ${\Rbar}_{k\ell}$ with the
Hamiltonian of Eq. (3.2) is far from trivial. It is
described in some detail in the Appendix. We present here
the results for the modulus squared, at $r=0$: we have
\begin{equation}
\left| {\Rbar}_{k\ell}(0) \right|^2 = \left[ 1 +2\,c_{\,\ell}(k)
\right] \,
\left| \widetilde{R}_{kl}(0) \right|^2 \ .\eqnum {3.9}
\end{equation}
Here $\widetilde{R}$ is evaluated with $H_{eff}^{(0)}$ so that
\begin{equation}
\left| \widetilde{R}_{k0}(0) \right|^2 =
{ \pi\,\CF \widetilde{\alpha}_{\hskip-1.3pt s}\,/v \over 1 -
e^{-\pi\CF\widetilde{\alpha}_{\hskip-1.3pt s}/v } }
\ \ \ , \eqnum {3.10.a}
\end{equation}
\begin{equation}
\left| 3\,\widetilde{R}^{\,'}_{k1}(0)/m\,v \right|^2 =
\left( 1 + {\CF^{\,2} \,\widetilde{\alpha}_{\hskip-1.3pt s}^{\,2} \over
4\,v^2}\right)\,
{ \pi\,\CF \widetilde{\alpha}_{\hskip-1.3pt s}\,/v \over 1 -
e^{-\pi\,\CF\,\widetilde{\alpha}_{\hskip-1.3pt s}\,/v } }
\ \ \ , \eqnum {3.10.b}
\end{equation}
$\widetilde{\alpha}_{\hskip-1.3pt s} = \widetilde{\alpha}_{\hskip-1.3pt
s}(\mu^2)$ given after Eq. (3.2).
The functions $c_{\,\ell}(k)$ are plotted in Figs. 1, 2 for
$\ell=0,1$, and displayed in detail in the Appendix. For $\ell=0,1$,
and small velocities,
\begin{eqnarray}
\nonumber
&& c_{\,0}(k) = {\BETAzero \, \alpha_{\hskip-1.pt s} \over 4\,\pi}
\,\left[ \ln {\mu\,a\over 2} - 1 - 2\,\Euler
+ {(ka)^2 \over 12} + {(ka)^4 \over 40} + \dots \right] \ ,\\
\nonumber
&& c_{\,1}(k) = {\BETAzero \, \alpha_{\hskip-1.pt s} \over 2\,\pi}
\Bigg[\;
\left( \ln {\mu a\over 2} - 2\,\gamma_E \right)\,
\left( -\,{3\over 2} + (ka)^2 - (ka)^4 \right)
- 2 + {49\over 24}\,(ka)^2  - {167 \over 80} \,(ka)^4 \\
\nonumber
&& \hskip5truecm + {1\over 2\,\pi} \left( ka - (ka)^3 + (ka)^5 \right)
+ \dots  \Bigg] \ ,\\
\eqnum {3.11}
&& a \equiv { 2\over m\,\CF \,\widetilde{\alpha}_{\hskip-1.3pt s}} \ \ \ .
\end{eqnarray}
A remarkable property of Eq. (3.11) is that, as
$k\rightarrow 0$ the relevant scale for
$\alpha_{\hskip-1.pt s}(\mu^2)$ is {\em{not}} $\mu \sim k$
(as one would naively guess, and as assumed for instance in Refs.
[8,9]), but $\mu \sim 2/a = m\,\CF\widetilde{\alpha}_{\hskip-1.3pt s}$:
the interaction {\em saturates}.

The full wave function is obtained by adding the (leading)
nonperturbative contributions. These may be deduced from the
the SVZ calculations$^{[1]}$. One finds
\begin{equation}
\left| R_{k\ell}(0) \right|^2 = \left( 1 + 2\, \delta_{k\ell}^{NP} \right)
\; \left| {\Rbar}_{k\ell}(0) \right|^2  \ \ . \eqnum {3.12}
\end{equation}
For $\ell=0$ one has
\begin{equation}
\delta_{k0}^{NP} = - { \pi\,\langle \alpha_{\hskip-1.pt s} \, G^2 \rangle
\over 192\, m^4\, v^6 } \ \ .
\eqnum {3.13}
\end{equation}
Just like for the bound states, the nonperturbative
correction blows up at threshold ($v\rightarrow 0$).
Clearly, the calculation ceases to be valid when $\left|
\delta^{NP} \right|$ is of the order of magnitude of
$|c_{\,\ell}|$, ${i.e.}$, for a critical velocity $v_{crit}$ such
that
\begin{equation}
v_{crit} \sim \left( { \pi^2 \langle \alpha_{\hskip-1.pt s} \,G^2\rangle
\over 192 \,\BETAzero \, m^4}
\right)^{1/6} \ \ \ . \eqnum {3.14}
\end{equation}

\section{Numerical results}
\label{sec:4}
\subsection{Production of $\;q{\qbar}\;$ above threshold.}
\label{subsec:4.1}
We now consider the quantities
\begin{eqnarray}
\nonumber
&& R_q^V(s) \ \equiv \ 12\,\pi\,Q_q^2\; {\rm{Im}}\;\Pi_V(s) \ \ , \\
\eqnum {4.1.a}
&& R_q^A(s) \ \equiv \ 12\,\pi\,Q_q^2\; {\rm{Im}}\;\Pi_A(s) \ \ .
\end{eqnarray}
$R_q^V$ is essentially the ratio $\;(e^+e^- \rightarrow
\gamma^* \rightarrow q{\qbar})/(e^+e^- \rightarrow \gamma^*
\rightarrow \mu^+\mu^-)$.  To show clearly the various
contributions, we will plot the zeroth and first order (in
$\alpha_{\hskip-1.pt s}$) expressions for the $R_q(s)$, cf.
Eq. (1.6):
\begin{eqnarray}
\nonumber
&& R_q^{V\;(0)}(s) = N_c \, Q_q^2\,{v(3-v^2) \over 2} \ \ , \\
\eqnum {4.2.a}
&& R_q^{V\;(1)}(s) \ {\mathop =_{v\to 0}} \ N_c \, Q_q^2\,
\left\{ {3\,\pi \over 4 } - {6\,v\over \pi} + {\pi\,v^2 \over 2} + \dots
\right\}\,\CF\,\alpha_{\hskip-1.pt s}(\mu^2) \ ,\\
\nonumber
&& R_q^{V\;(1)}(s) \ {\mathop =_{v\to 1}} \ N_c \, Q_q^2\,
\left\{ {3\,\over 4 } + {9\over 2}(1-v) + \left( {9\over 2} \ln
{2\over 1-v} - {3\over 8} \right) (1-v)^2 + \dots
\right\}\,{\CF\,\alpha_{\hskip-1.pt s}(\mu^2)\over \pi} \ ;\\
\nonumber
 \\
\nonumber
&& R_q^{A\;(0)}(s) = N_c \, Q_q^2\,v^3 \ \ , \\
\eqnum {4.2.b}
&& R_q^{A\;(1)}(s) = N_c \, Q_q^2\,
\left\{ \pi\,\CF\,\alpha_{\hskip-1.pt s}(\mu^2)\,v^2 + \dots
\right\} \ .
\end{eqnarray}
The expression for $R_q^{V\;(1)}$ is actually known
exactly$^{[10]}$, but Eq. (4.2.a) is accurate enough for our
purposes. Only the leading term (in $v$) in $R_q^{A\;(1)}$
is known, and it is reported in Eq.  (4.2.b).
It is worth noting that the Shwinger interpolation$^{[10]}$ to
$R_q^{V\;(1)}$, used by some authors, is {\em not} accurate enough
for our purposes. In fact, it reproduces correctly the values of
$R_q^{V\;(1)}(s)$ for $v=0,\ \, v=1\,$, but not the {\em derivatives},
$i.e.$, the terms in $v,\ v^2$; $(1-v),\ (1-v)^2\,$. For this reason
we take the full (4.2.a) (see below, e.g. in (4.3.a)).

Together with these perturbative evaluations we also plot
the ``exact'' expressions obtained from Eqs. (3.9--3.13):
\begin{eqnarray}
\nonumber
R_q^{V}(s) \ {\mathop =_{ v\to 0}} \  N_c \, Q_q^2\,&&\left\{
{v(3-v^2) \over 2} + \left(  - {6\,v\over \pi} + {3\,\pi\,v^2 \over 4 }\right)
\,\CF\,\widetilde{\alpha}_{\hskip-1.3pt s} \right\}
\left( 1 - {\pi \langle \alpha_{\hskip-1.pt s} \,G^2\rangle   \over 192 \,
m^4\,v^6} \right) \\
\eqnum {4.3.a}
&& \times \big[ 1+2\,c_{\,0}(k) \big] \,
{ \pi\,\CF \widetilde{\alpha}_{\hskip-1.3pt s} /v \over 1 -
e^{-\pi\CF\widetilde{\alpha}_{\hskip-1.3pt s}/v } }
\hskip0.8truecm , \hskip0.8truecm \widetilde{\alpha}_{\hskip-1.3pt s} =
\widetilde{\alpha}_{\hskip-1.3pt s} (\mu^2)
\ ;
\end{eqnarray}
we have included the known ${\cal{O}}(\alpha_{\hskip-1.3pt
s} v,\ \alpha_{\hskip-1.pt s} v^2)$ corrections, cf. Eq.
(4.2.a). For $R_q^A$,
\begin{equation}
R_q^{A}(s) \ {\mathop =_{v\to 0}}\
N_c \, Q_q^2\,v^3\,(1+2\,\delta_{k1}^{NP} )\,
\big[ 1+2\,c_{\,1}(k) \big] \,
\left( 1 + { \CF^{\,2} \,\widetilde{\alpha}_{\hskip-1.3pt s}^{\,2} \over 4\,v^2
} \right) \,
{ \pi\,\CF \widetilde{\alpha}_{\hskip-1.3pt s} /v \over 1 -
e^{-\pi\CF\widetilde{\alpha}_{\hskip-1.3pt s}/v } }\ .
 \eqnum {4.3.b}
\end{equation}
The quantities $\delta_{k1}^{NP}$ may be found in Ref. [11].
The calculation of $R_q^{V}$ will be made for $q=c,b\;$;
$R_q^{A}$ will also be evaluated for $q=t$.

Before presenting the results, a few words have to be said
about the parameters entering Eqs. (4.2, 4.3). For the
masses we will take the values given in Eqs. (3.4, 3.5).
For $\alpha_{\hskip-1.pt s}(\mu^2)$, the ``natural'' scale
is $\mu\sim m\,\CF\,\widetilde{\alpha}_{\hskip-1.3pt s}$, as
shown in Eqs. (3.11).  Nevertheless, for $c,\ b$ we will
choose $\mu = 2\,m_q$ (but see subsection B for a discussion
of this).  For $t{\tbar}$ production, and since the choice
of $\mu$ is less relevant now, we will take $\mu = M_Z\;$;
then we may use directly the value of $\alpha_{\hskip-1.3pt
s}$ deduced from the $Z$ decays. Thus, we take
\begin{equation}
\alpha_{\hskip-1.pt s}(M_Z) = 0.119 \pm 0.003\ , \eqnum {4.4.a}
\end{equation}
for $t{\tbar}$; and for $b{\bbar}$ and $c{\cbar}$,
\begin{equation}
\alpha_{\hskip-1.pt s}(\mu^2) = { 4\,\pi \over \BETAzero\,\ln\,\mu^2/\Lambda^2
}
\left\{\;
 1 - { \BETAone \, \ln\,\ln\,\mu^2/\Lambda^2 \over
\BETAzero^2\,\ln\,\mu^2/\Lambda^2 }
\;\right\} \ ,
\eqnum {4.4.b}
\end{equation}
\begin{displaymath}
\mu =2\,m_c,\ 2\,m_b\ \ ; \ \ \ \Lambda=200 ^{+80}_{-60}\ MeV\ \ , \ \ \
\BETAone = 102 - 38\,n_f/3\ .
\end{displaymath}
Finally, for $\langle \alpha_{\hskip-1.pt s} G^2\rangle $
(which does not play a very important role in our
evaluations) we choose the standard value$^{[12]}$
\begin{equation}
\langle \alpha_{\hskip-1.pt s} G^2\rangle  = 0.042 \pm 0.020\ GeV^4 \ .
\eqnum {4.5}
\end{equation}

As stated several times, our calculations are valid in the
nonrelativistic regime, $i.e.$, to corrections
${\cal{O}}(v^2)$. For the numerical evaluations we will take
$v<v_{\rm{Max}}\;,\ v_{\rm{Max}}=1/2\; .$

\subsection{Comparison with experiment: $\;c{\cbar}\;$ and $\;b{\bbar}$.}
\label{subsec:4.2}
The prediction of our calculation (4.3.a) for $\;c{\cbar}$
is shown in Fig. 3. In order to display the dependence of
the calculation on the choice of the renormalization scale,
$\mu$, we present these $R_c^V$ for two choices of $\mu$:
electing $\mu$ such that the radiative corrections vanish,
$i.e.$, such that $c_0=0$, or fixing $\mu=2\,m_c$. There is
little difference between both choices; here we favour this
last choice because it ties with the election at higher
energies ($s^{1/2}\gg m_c$) where one takes $\mu=s^{1/2}$.
Also shown are the results of a parton model ($R_c^{V
\;(0)}$) and parton model plus order $\alpha_{\hskip-1.3pt
s}(m_c)$ correction (the last denoted by $R_c^{V\;(1)}$). It
is seen that there is a partial cancellation of the Fermi
factor,
\begin{displaymath}
{ \pi\,\CF \widetilde{\alpha}_{\hskip-1.3pt s} /v \over 1 -
e^{-\pi\CF\widetilde{\alpha}_{\hskip-1.3pt s}/v } }\ \ ,
\end{displaymath}
and the radiative correction, $2\,c_0(k)$ (c.f. Eq. (4.3.a))
in such a way that $R_q^V$ does not differ much from
$R_c^{V\;(0)+(1)}$.  The nonperturbative contribution is
only important right at threshold.

Comparison with experiment is shown in Fig. 4. Although the
quality of this is not enchanting as a fit, a few things
must be said in its favour.  First, the theoretical curve
runs, more or less, through the middle of the experimental
points. It is clear that the full $R_c^{V\;({\rm{exact}})}$
is more centered than the purely partonic $R_c^{V\;(0)}$ or
what we would have obtained neglecting the radiative
correction, $R_c^{V\;({\rm{no\ correc.}})}$ (in this last
case, the improvement is slight). {\em{On the average}},
$R_c^{V\;({\rm{exact}})}$ represents a good mean of
experiment, which is an interesting fact, particularly since
part of the dispersion of the experimental points is
doubtlessly due to error fluctuations.

Fig. 5 shows the comparison with experiment for $b{\bbar}$.  The
conclusions are similar to those for $c\cbar$:
$R_b^{V\;({\rm{exact}})}$ improves $R_b^{V\;(0)}$ and represents a
good average of the experimental points, which are now scantier.

\subsection{$\;t{\tbar}$ production.}
\label{subsec:4.3}
We plot in Figs. 6, 7 the quantities $R^V$, $R^A$ relevant
to production of $t{\tbar}$ by a vector, axial current
respectively. It is not possible to use them directly to
predict experimental output because our formulas do not take
into account the width of $t$. This can be done with the
standard methods$^{[13]}$, and we will present the details
separately.

\section{Threshold effects on ``low energy'' correlators.}
\label{sec:5}
Consider a correlator, $\Pi(t)$. It is possible to prove
quite generally that it verifies the relation
({\em{dispersion relation}})
\begin{equation}
\Pi(t) - \Pi(0)  = {t \over \pi} \int^{\infty} ds \;
{ {\rm {Im}}\;\Pi(s) \over s(s-t) } \ \ . \eqnum {5.1}
\end{equation}
By expanding in a power series in $\alpha_{\hskip-1.pt s}$
it follows that Eq. (5.1) is also verified order by order in
perturbation theory:
\begin{equation}
\Pi^{(n)}(t) - \Pi^{(n)}(0)  = {t \over \pi} \int^{\infty} ds \;
{ {\rm {Im}}\;\Pi^{(n)}(s) \over s(s-t) } \ \ . \eqnum {5.2}
\end{equation}
When used in this last form, the dispersion relation is
little more than a calculational device which may simplify
the evaluation of $\Pi^{(n)}$ as, generally speaking,
${\rm{Im}}\,\Pi^{(n)}$ is easier to calculate than
${\rm{Re}}\,\Pi^{(n)}$. This method is followed e.g.  in
Refs. [10] $\;$ for $\Pi_V^{(1)}$. When used in the form
(5.1), however, a dispersion relation may yield new
knowledge.  This happens when experimental data may be used
as input for ${\rm{Im}}\,\Pi$, thus obtaining values of
$\Pi$ in regions inaccessible to theory, as is the case for
evaluations of the hadronic corrections to the anomalous
magnetic moment of the muon. Another situation, which is the
one we will encounter here, is when nonperturbative methods
are employed to evaluate ${\rm{Im}}\,\Pi$ (or parts
thereof).

Let us define the {\em{threshold effect}} contributions to
$\Pi(t) - \Pi(0)$, to be denoted by
$\Delta_{{\rm{th}}}^{(n)}(t)$, to be the quantity
\begin{eqnarray}
\eqnum {5.3}
&& \Delta_{{\rm{th}}}^{(n)}(t) = {t\over \pi} \int^{s_{\rm{M}}}
ds\; { f^{(n)} (s) \over s(s-t) } \ \, \\
\nonumber
&& f^{(n)} (s) \equiv
{\rm{Im}}\,\Pi(s) - \displaystyle  \sum_{n'=0}^{n} {\rm{Im}}\,\Pi^{(n')}(s)
\end{eqnarray}
That is to say, $\Delta_{{\rm{th}}}^{(n)}$ incorporates the
exact contribution to $\Pi(t) - \Pi(0)$ from a region around
threshold, up to an energy $s_{\rm{M}}$, but subtracting the
first $n$ terms in perturbation theory. Clearly, one has
\begin{equation}
\Pi(t) - \Pi(0) = \sum_{n'=0}^{n} \left\{
\Pi^{(n')}(t) - \Pi^{(n')}(0) \right\} + \Delta_{{\rm{th}}}^{(n)}(t)
+ \Delta_{{\rm{h.e}}}^{(n)} \ \ ,\eqnum {5.4}
\end{equation}
where $\Delta_{{\rm{h.e}}}^{(n)}$ would be
\begin{equation}
\Delta_{{\rm{h.e}}}^{(n)}(t) = {t \over \pi}
\int_{s_{\rm{M}}}^{\infty}ds \;
{ {\rm{Im}}\,\Pi(s) - \displaystyle  \sum_{n'=0}^{n} {\rm{Im}}\,\Pi^{(n')}(s)
\over s(s-t) } \ \ . \eqnum {5.5}
\end{equation}
Expression (5.4) is then useful when we can argue that
$\Delta_{{\rm{h.e}}}^{(n)}$ is small with respect to
$\Delta_{{\rm{th}}}^{(n)}$, or (as happens for $\Pi_V$) when
the difference
\begin{displaymath}
{\rm{Im}}\,\Pi(s) - \displaystyle  \sum_{n'=0}^{n} {\rm{Im}}\,\Pi^{(n')}(s)
\end{displaymath}
may be calculated for $s>s_{\rm{M}}$.

It should be clear that $\Delta_{{\rm{th}}}^{(n)} +
\Delta_{{\rm{h.e}}}^{(n)}$ depends on $s_{\rm{M}}$;
only if {\em{exact}} evaluations were used that
an exact match (and thus cancellation) of
the dependence of each of $\Delta_{{\rm{th}}}^{(n)}$,
$\Delta_{{\rm{h.e}}}^{(n)}$ on $s_{{\rm{M}}}$ would occur.
In favorable cases we expect, however, that the residual
dependence would be slight.


\subsection{The axial correlator.}
\label{subsec:5.1}
We do not have any information on
$\Delta_{{\rm{h.e}}}^{(n)}$ for the axial correlator, but we
will give results on $\Delta_{{\rm{th}}\;A}^{(2)}$ for
completeness. Separating the contribution from the bound
states and the piece above threshold, we write
\begin{equation}
\Delta_{{\rm{th}}\;A}^{(n)} = \Delta_{{\rm{pole}}\;A}^{(n)}
+ \Delta_{{\rm{a.t.}}\;A}^{(n)}  \ \ . \eqnum {5.6}
\end{equation}
Using Eq. (2.4.b) for $\Delta_{{\rm{pole}}\;A}^{(2)}$, and Eqs. (4.1),
(4.2.b), (4.3.b) for $\Delta_{{\rm{a.t.}}\;A}^{(2)}$, we obtain
\begin{equation}
\Delta_{{\rm{pole}}\;A}^{(1)} (t) = {3\,N_c\,t \over 2\,m^2\,\pi}
\sum_{N} {1 \over M_{{\rm{N1}}}^3 } \;
	 {1 \over M_{{\rm{N1}}}^2 - t } \; \big| R_{{\rm{N1}}}^{\,'}(0) \big|^2
\ \ ,
\eqnum {5.7}
\end{equation}
where the sum runs over the $\ell=1$ bound states, and
\begin{eqnarray}
\nonumber
\Delta_{{\rm{a.t.}}\;A}^{(1)} (t) =&& {N_c\,t \over 12\,\pi^2}
 \int_{s_{{\rm{th}}}}^{s_{{\rm{M}}}} ds \; {1 \over s\,(s-t) } \\
\eqnum {5.8}
&& \times \Bigg\{ v^3 \left( 1-2\,\delta_{k1}^{NP} \right)
\left[ 1+2\,c_{1}(k) \right] \,
\left( 1- {\CF^2 \,\widetilde{\alpha}_{\hskip-1.3pt s}^2 \over 4 \, v^2}
\right)
\; { \pi \CF \widetilde{\alpha}_{\hskip-1.3pt s}/v \over
	1 - e^{ -\pi \CF \widetilde{\alpha}_{\hskip-1.3pt s}/v } } \\
\nonumber
&& \ \ \ \ - v^3 - \pi \CF \alpha_{\hskip-1.pt s} v^2 \Bigg\}
\ \ , \ \ \ \ k\equiv mv \ \ , \ \ v \equiv \sqrt{1-4\,m^2/s} \ .
\end{eqnarray}
For the case of $\,t{\tbar}\,$ production, the
nonperturbative corrections $\,\delta_{k1}^{NP}\,$ are
negligible, so in this particular case we may approximate
the threshold $\,s_{{\rm{th}}}\,$ by $\;4\,m_t^2\;$.
Moreover, the spectrum of bound states is Coulombic with a
very good approximation up to $\,N\sim 4\,$; since bound
states with $\,N>4\,$ contribute very little to Eq. (5.7)
anyway, we may take the spectrum to be Coulombic all the
way.  Thus, we may rewrite Eqs.  (5.7), (5.8) as
\begin{equation}
\Delta_{{\rm{pole}}\;A}^{(1)} (t) = {N_c\,t \over m^2\,\pi}
\sum_{N=2}^{\infty}
{1 \over M_{{\rm{N1}}}^3 \left( M_{{\rm{N1}}}^2 - t \right) }\;
	 {N+1 \over N^3(N-1)\,a(N,1)^5} \ \ ,
\eqnum {5.9.a}
\end{equation}
$a(N,1)\,$ given in Eq. (3.7), and the $\,M_{{\rm{N}}}\,$
are as in Eq. (3.6) (neglecting the $\,NP\,$ piece),
\begin{equation}
M_{{\rm{N1}}} = 2\,m
\left\{
1 - { \CF^2 \widetilde{\alpha}_{\hskip-1.3pt s}(\mu^2)^2 \over 8\,N^2 }
  - { \CF \BETAzero \alpha_{\hskip-1.pt s}^2 \widetilde{\alpha}_{\hskip-1.3pt
s} \over 8\,\pi\,N^2 }
\left[ \ln{\mu N \over m\CF\widetilde{\alpha}_{\hskip-1.3pt s} } + \psi(N+2)
\right] \;
\right\} \ . \eqnum {5.9.b}
\end{equation}
Likewise,
\begin{eqnarray}
\nonumber
\Delta_{{\rm{a.t.}}\;A}^{(1)} &&(t) = {N_c\,t \over 12\,\pi^2}
 \int_{4\,m^2}^{s_{{\rm{M}}}} ds \; {1 \over s\,(s-t) } \\
&& \times \Bigg\{ v^3
\left[ 1+2\,c_{1}(k) \right] \,
\left( 1- {\CF^2 \,\widetilde{\alpha}_{\hskip-1.3pt s}^2 \over 4 \, v^2}
\right)
\; { \pi \CF \widetilde{\alpha}_{\hskip-1.3pt s}/v \over
	1 - e^{ -\pi \CF \widetilde{\alpha}_{\hskip-1.3pt s}/v } }
 - v^3 - \pi \CF \alpha_{\hskip-1.pt s} v^2 \Bigg\}  \ .
\eqnum {5.10}
\end{eqnarray}

The practical interest of this is fairly limited so long as
we have no reliable estimate of
$\,\Delta_{{\rm{h.e}}\;A}^{(1)}\,$.  We then turn to the
vector correlator, for which such an estimate exists.

\subsection{The vector correlator.}
\label{subsec:5.2}

The calculation is very much like for the axial case, except
that we use Eq. (2.4.a) for the poles, and Eqs. (4.1),
(4.2.a), (4.3.a) above threshold. We find, for
$\,t{\tbar}\,$ and neglecting the gluon condensate
contribution,
\begin{equation}
\Delta_{{\rm{pole}}\;V}^{(1)} (t) = {4\,N_c\,t \over \pi}
\sum_{N=1}^{\infty}
{1 \over M_{{\rm{N0}}}^3 \left( M_{{\rm{N0}}}^2 - t \right) }\;
	 {1 \over N^3\;a(N,0)^3} \ \ ,
\eqnum {5.11}
\end{equation}
and now
\begin{equation}
M_{{\rm{N0}}} = 2\,m
\left\{
1 - { \CF^2 \widetilde{\alpha}_{\hskip-1.3pt s}(\mu^2)^2 \over 8\,N^2 }
  - { \CF \BETAzero \alpha_{\hskip-1.pt s}^2 \widetilde{\alpha}_{\hskip-1.3pt
s} \over 8\,\pi\,N^2 }
\left[ \ln{\mu \over m\CF\widetilde{\alpha}_{\hskip-1.3pt s} } + \psi(N+1)
\right] \;
\right\} \ . \eqnum {5.12}
\end{equation}

We split the region above thresholds into two parts, a low
energy ($l.e.$) and a high energy ($h.e.$) part, according
to $v<1/2$ or $v>1/2$. Note that $v=1/2$ occurs for $s=16\,m^2/3$.
Later we will discuss the joining of the two regions.

At low energy we have,
\begin{equation}
\Delta_{{\rm{l.e.}}\;V}^{(1)} (t) = {N_c\,t \over 12\,\pi^2}
 \int_{4\,m^2}^{16\,m^2/3} ds \; {f_{{\rm{l.e.}}}^{(1)}(s) \over
s\,(s-t) } \ \ ,\eqnum {5.13.a}
\end{equation}
and, from Eqs. (4.3.a), (4.2.a)
\begin{eqnarray}
\nonumber
f_{{\rm{l.e.}}}^{(1)}(s) && \equiv R_t^V - R_t^{V\;(0)+(1)} \\
\nonumber
&& = { \pi \CF \widetilde{\alpha}_{\hskip-1.3pt s}(\mu^2)/v \over
	1 - e^{ -\pi \CF \widetilde{\alpha}_{\hskip-1.3pt s}(\mu^2)/v } } \left\{
{v(3-v^2)
\over 2 } + \left( -{6\,v\over \pi} + {3\,\pi v^2 \over 4}
\right)\CF\widetilde{\alpha}_{\hskip-1.3pt s}(\mu) \right\}\; \left[
1+2\,c_{0}(k)
\right] \; \\
\eqnum {5.13.b}
&& \ \ \  - {v(3-v^2) \over 2}
- \left( {3\,\pi \over 4} - {6\,v\over \pi} + {\pi v^2 \over 2} \right)
\CF \alpha_{\hskip-1.pt s}(\mu^2) \; \ , \\
\nonumber
&& \ \ \ k = mv \ \ , \ \ v=\sqrt{1-4\,m^2/s} \ \ , \ \ \mu=M_Z \ .
\end{eqnarray}
For high energy there exist evaluations$^{[14]}$ correct to
errors $(1-v)^2\;\alpha_{\hskip-1.pt s}^2\;$,
$\alpha_{\hskip-1.pt s}^4$. Subtracting from this
the order $\alpha_{\hskip-1.pt s}$ piece, we may write the result
as
\begin{equation}
f_{{\rm{h.e.}}}^{(1)} \equiv {\fbar}_{{\rm{h.e.}}}
- f_{{\rm{h.e.}}}^{(0+1)}\ \ ,
\eqnum {5.14.a}
\end{equation}
where $f_{{\rm{h.e.}}}^{(0+1)}$ is the piece of order
$(0+1)$ in $\alpha_{\hskip-1.pt s}$,
\begin{eqnarray}
\nonumber
f_{{\rm{h.e.}}}^{(0+1)} =&& -{3\over 2}(1-v)^2 + {1\over 2}(1-v)^3\\
\eqnum {5.14.b}
&& + \left[ {9\over 2} (1-v) + {9\over 2} \left( \ln{2\over 1-v} -{3\over 8}
\right)(1-v)^2 \right] {\CF \alpha_{\hskip-1.pt s}\over \pi} \ ,
\end{eqnarray}
and$^{[14]}$,
\begin{eqnarray}
\nonumber
{\fbar}_{{\rm{h.e.}}} &&= -{3\over 2}(1-{\vbar})^2
+ {1\over 2}(1-{\vbar})^3
+ \left[ {9\over 2} (1-{\vbar})
+ {9\over 2} \left( \ln{2\over 1-{\vbar}} -{3\over 8}
\right)(1-{\vbar})^2 \right] {\CF \alpha_{\hskip-1.pt s}\over \pi} \\
\eqnum {5.14.c}
&& \
+ r_2 \, \left( {\alpha_{\hskip-1.pt s}(s) \over \pi }\right)^2
+ \widetilde{r}_3 \, \left( {\alpha_{\hskip-1.pt s} \over \pi }\right)^3
+ {9\over 2}(1-v) {\CF}
	\left[
		8.7 \,\left({\alpha_{\hskip-1.pt s}(s) \over \pi } \right)^2
		+ 45.3 \,\left({\alpha_{\hskip-1.pt s}(s) \over \pi }\right)^3
	\right]\;.
\end{eqnarray}
Here,
\begin{eqnarray}
\nonumber
&& {\vbar} = \sqrt{ 1 - {\mbar}^2(s)/s } \ \ \ \ , \ \ \ \ \
   r_2 = 1.986 -0.115\,n_f \ \ \ ,\\
\nonumber
&& \widetilde{r}_3 = -6.637 -1.2\,n_f -0.005\,n_f^2
- 1.24 \left( \sum_f Q_f \right)^2 \ \ \ ,
\end{eqnarray}
and the running mass ${\mbar}(\mu)$ is given in terms of $m$ by
\begin{eqnarray}
\nonumber
&& {\mbar} = m\,\left[ { \alpha_{\hskip-1.pt s}(\mu) \over \alpha_{\hskip-1.pt
s}(m) } \right]^{d_m}\;
\left\{
  1 + { A\,\alpha_{\hskip-1.pt s}(\mu) - (\CF-A)\alpha_{\hskip-1.pt s}(m) \over
\pi}
\right\} \ \ ,\\
\nonumber
&& A = { \BETAone \gamma_0 - \BETAzero \gamma_1 \over \BETAzero^2}
\ \ \ , \ \ \ d_m = - {\gamma_0 \over \BETAzero } \ \ ,\\
\nonumber
&& \gamma_0 = - 3\,\CF \ \ \ , \ \ \ \gamma_1 = - {3\,\CF^2 \over 2}
- {97\,\CF\CA \over 6} + { 5\,\CF n_f \over 3}
\end{eqnarray}
Finally,
\begin{equation}
\Delta_{{\rm{h.e.}}\;V}^{(1)} (t) = {N_c\,t \over 12\,\pi^2}
 \int_{16\,m^2/3}^{\infty} ds \; {f_{{\rm{h.e.}}}^{(1)}(s) \over
s\,(s-t) } \ \ . \eqnum {5.15}
\end{equation}
$\Delta_{{\rm{pole}}\;V}^{(1)}$, $\Delta_{{\rm{l.e.}}\;V}^{(1)}$, and
$\Delta_{{\rm{h.e.}}\;V}^{(1)}$ should be compared with the
$(0+1)$--order direct calculation of $\Pi$: for small $t$,
\begin{eqnarray}
\nonumber
\Pi_1(t) && \equiv \sum_{n=0}^{1} \left\{ \Pi^{(n)}(t) - \Pi^{(n)}(0)
\right\} \\
\eqnum {5.16}
&& = {N_c \,t \over 12\,\pi^2}\, {1 \over 5\,m^2} \Bigg\{ 1 - {3\over
28}\,{t\over m^2} + {205\over 54}\, {\CF \alpha_{\hskip-1.3pts}
(\mu^2) \over \pi} + \dots
\Bigg\}
\end{eqnarray}
Numerically,
\begin{equation}
{3\,M_Z^2\over 28\,m^2} = 3.2 \times 10^{-2} \ \ \ \ \ \ \ ;
\ \ \ \ \ \ \ {205\over 54}\, {\CF \alpha_{\hskip-1.pt s}(M_Z^2)
\over \pi} = 0.19 \pm 0.005\ \ .
\eqnum {5.17}
\end{equation}
The error in the above equation is due to the
experimental error in $\alpha_s(M_Z)$.

As for the $\Delta^{(1)}$, we write, for the various
contributions,
\begin{equation}
\Delta^{(1)}(t) = {N_c \,t \over 12\,\pi^2}\, {1 \over 5\,m^2}\;
\widehat{\Delta}^{(1)}(t) \ \ .\eqnum {5.18}
\end{equation}
In this way we may compare directly with Eq. (5.17). Then,
\begin{eqnarray}
\nonumber
&& \widehat{\Delta}^{(1)}_{{\rm{pole}}\;V}(0) = 2.33\times 10^{-2} \\
\eqnum {5.19}
&& \widehat{\Delta}^{(1)}_{{\rm{l.e.}}\;V}(0) =  1.56\times 10^{-2} \\
\nonumber
&& \widehat{\Delta}^{(1)}_{{\rm{h.e.}}\;V}(0) = 2.15\times 10^{-2}
\end{eqnarray}
The dependence of the $\widehat{\Delta}$ on $t$ is very
slight, up to $t=M_Z^2$, where we have
\begin{eqnarray}
\nonumber
&& \widehat{\Delta}^{(1)}_{{\rm{pole}}\;V}(M_Z^2) = 2.52\times 10^{-2}\\
\eqnum {5.20}
&& \widehat{\Delta}^{(1)}_{{\rm{l.e.}}\;V}(M_Z^2) = 1.67\times 10^{-2}\\
\nonumber
&& \widehat{\Delta}^{(1)}_{{\rm{h.e.}}\;V}(M_Z^2) = 2.24\times 10^{-2}
\end{eqnarray}

The largest error of the $\Delta$ is due to the error in the
mass of the $t\,$ quark, about which little can be done at
present.  Another source of error is due to extrapolations:
we have used Eq. (5.13) for $f_{{\rm{l.e.}}}$ up to $v=1/2$,
and Eq. (5.14) for $f_{{\rm{h.e.}}}$ down to same value of
$v$. We can smooth this rather crude joining of l.e. and
h.e. regions by defining
\begin{equation}
f_1 \equiv (1-v)\,f_{{\rm{l.e.}}} + v \, f_{{\rm{h.e.}}} \eqnum {5.21.a}
\end{equation}
and integrating this $f_1$ over all the interval, from
$4\,m^2$ to infinity. Eqs. (5.19) are not changed
substantially; we obtain now, with self-explanatory
notation,
\begin{equation}
\widehat{\Delta}^{(1)}_{{\rm{l.e.+h.e.}}\;V\,,\,1}(0) = 4.1 \times 10^{-2}
\ \ .
\eqnum {5.21.b}
\end{equation}
Another possibility is to write
\begin{equation}
f_3 \equiv (1-v^3)\,f_{{\rm{l.e.}}} + v^3 \, f_{{\rm{h.e.}}} \ \ ,
\eqnum {5.22.a}
\end{equation}
which respects the terms in $v^0,\ v,\ v^2$ which are known
at low energy.  Then,
\begin{equation}
\widehat{\Delta}^{(1)}_{{\rm{l.e.+h.e.}}\;V\,,\,3}(0) = 3.3 \times 10^{-2}
\ \ .
\eqnum {5.22.b}
\end{equation}
We consider Eqs. (5.21) to be the more reasonable estimate,
and take its difference with Eqs. (5.19), (5.22) to be a
measure of the systematic theoretical errors in our
calculation\footnote{To which one should add errors due to
experimental errors in $m_t$,
$\alpha_{\hskip-1.pt s}(M_Z)$.}.

Thus we finally get
\begin{eqnarray}
\nonumber
\widehat{\Delta}^{(1)}_{{\rm{all}}\;V}(0) &&=
\widehat{\Delta}^{(1)}_{{\rm{pole}}\;V}(0) +
\widehat{\Delta}^{(1)}_{{\rm{l.e.}}\;V}(0) +
\widehat{\Delta}^{(1)}_{{\rm{h.e.}}\;V}(0)\\
\eqnum {5.23}
&& = (\;4.1 \pm 0.8\;) \times 10^{-2} \ \ .
\end{eqnarray}
This may be compared with the evaluations of Refs. [8,9,15], which are
clearly improved by our results. In particular, the unstabilities
noted by Gonzalez-Garcia et al.$^{[15]}$ disappear almost completely.
This is due to our use of information from the {\em high energy}
region (which eliminates the uncertainties due to the dependence on an
energy cut-off), and inclusion of the radiative corrections, which
reduce drastically the arbitrariness of the choice of the scale $\mu$
of $\alpha_{\hskip-1.pt s}(\mu)$.

Although the evaluation is reasonably reliable, it should be
clear that the effect is small in the sense that
$\widehat{\Delta}_{{\rm{all}}\;V}$ as given by Eq. (5.23) is
smaller than the perturbatively known piece, Eq. (5.17), by
a factor of about 5.
%

\section{Acknowledgements}
The authors are grateful to S.~Titard for useful
discussions. The collaboration of G.~Lopez~Castro in the
early stages of this work should be mentioned with
gratitude.

\appendix
\section{Wave functions in the continuum}
\label{app:1}

We present here some of the technicalities used to solve the
radial wave equation in the continuum. Let us recall that we
are only interested in obtaining the value of the wave
function at the origin for $\ell=0$, and its derivative at
the origin for $\ell=1$.  This simplifies considerably the
calculation.  The main results of this Appendix were given
in Eq. (3.11).

To find the radial solutions ${\Rbar}_{k\ell}(r)$ of the
Hamiltonian $H$, as given in Eq. (3.2), we treat $H_1$ to
first order, and use the method of variation of constants.
In other words, we want to solve
\begin{eqnarray}
\nonumber
&& {\Rbar}_{k\ell}^{\,''}(r) + {2\over r} \,
{\Rbar}_{k\ell}^{\,'}(r) +
\left( - {\ell(\ell+1) \over r^2} + k^2 + {2\over ar} + {2\,\lambda \;
\ln\, r\mu \over a\,r} \right)\,{\Rbar}_{k\ell} (r) = 0 \ \ ,\\
&& k = mv \ , \ \ \ a = { 2 \over m \CF \widetilde{\alpha}_{\hskip-1.3pt
s}(\mu^2) } \ , \ \ \
\lambda = { m\,a\,\CF \BETAzero \alpha_{\hskip-1.pt s}^2 \over 4\pi}\ ,
\eqnum {A.1}
\end{eqnarray}
taking $\lambda$ infinitesimal. For $\lambda=0$, the {\em{normalized
solution}} (see Eq. (2.8))  which is regular at $r=0$ is
\begin{equation}
\widetilde{R}_{k\ell}(r) = e^{i\delta} \; { \left|\,\Gamma(A_{\ell})\,\right|
\over \Gamma(2\ell+2) } \;
e^{\pi/2ka} \; e^{-ikr} \; (2kr)^{\ell} \; M(A_{\ell},2{\ell}+2,2ikr) \ \ ,
\eqnum {A.2}
\end{equation}
where
\begin{equation}
A_{\ell}\equiv {\ell}+1+{i\over ka} \ \ , \eqnum {A.3}
\end{equation}
$M$ is the Kummer function, and $\delta$ an arbitrary phase,
which we choose to be zero.

Let us now discuss the case $\ell=0$. The general solution
of Eq. (A.1) is
\begin{eqnarray*}
&& {\Rbar}_{k0} = \widetilde{R}_{k0} + \delta R_{k0} \ \ , \\
&& \delta R_{k0} = e^{\pi/2 ka} \left|\,\Gamma(A_0) \,\right| \,
e^{-ikr} \,Y(r) \ \ ,\\
&& Y(r) = c_{\,0}(k) \; M(A_0,2,2ikr)
+ 2ikr \bigg[ M(A_0,2,2ikr) \int_0^r d\rho \,\varphi_1 \\
&& \hskip6.2truecm + U(A_0,2,2ikr) \int_0^r d\rho \,\varphi_2 \bigg] \ ,
\end{eqnarray*}
where
\begin{displaymath}
\varphi_1 = UX/W \ \ , \ \ \varphi_2 = -MX/W\ \ , \ \
X = {i\lambda \over ka}\, { \ln \,r\mu \over 2ikr } \;
M(A_0,2,2ikr) \ \ ,
\end{displaymath}
$U$ is the Kummer function which is singular at $r=0$,
and $W$ is the Wronskian of $U$ and $M$,
\begin{displaymath}
W^{-1} = \Gamma(A_0)\, e^{-2ikr} \,(2ikr)^2 \ \ .
\end{displaymath}
A term in $Y$ of the form $\ C(k)\, U \ $ is excluded by the
condition of regularity at $r=0$. The constant $c_{\,0}$
must be determined from the condition of normalization of
${\Rbar}_{k0}$, which is equivalent here to the
orthogonality of $\widetilde{R}_{k0} $ and $\delta\,R_{k0}$:
\begin{displaymath}
 \int_0^{\infty} dr\; r^2\,\widetilde{R}_{k0}^{\;*} \;
\delta R_{k0} \ =\  0 \ \ \ .
\end{displaymath}
After matching infinities, the value of $c_{\,0}(k)$ follows:
\begin{eqnarray}
\nonumber
c_{\,0}(k) && = - 2ik \int_0^{\infty} dr \, r^2 \,
\left\{ \;
\varphi_1 - {\Gamma(A_0^*) \, e^{\pi/ka}\, \over 2} \;
\varphi_2 \; \right\} \\
\eqnum {A.4}
&& = \ {4\,ik\,\lambda \, \Gamma(A_0) \over a}  \int_0^{\infty} dr \,
(r \ln \,\mu r ) \,
\left\{ \;
M^* U + {\Gamma(A_0^*) e^{\pi/ka} \over 2} \, |M|^2
\;\right\}\ .
\end{eqnarray}
The arguments in both Kummer functions $M, \ U$ are the
same, e.g., $M = M(A_0,2,2ikr)$.

As we stated earlier, we are only interested in $R_{k0}(0)$.
We then have,
\begin{displaymath}
R_{k0}(0) = e^{-\pi/2ka} \, \Gamma(A_0^*) \; \big[\; 1+c_{\,0}(k)\; \big]
\ \ .
\end{displaymath}
To complete the calculation it only remains to evaluate
$c_{\,0}(k)$.  We will come back to this later.

Repeating the above analysis for the case $\ell=1$ gives:
\begin{equation}
 {{\Rbar}_{k1} \over r }\Bigg|_{r=0} =  \big[\;1+c_{\,1}(k) \;\big] \
{\widetilde{R}_{k1} \over r }\Bigg|_{r=0} \ \ ,\eqnum {A.5}
\end{equation}
where
\begin{equation}
 c_{\,1}(k) =
\lambda\,{ (2ik)^3 \; \Gamma(A_1) \over 3\,a }
\int_{0}^{\infty} dr \, r^3 \, \ln\left(\mu r\right) \; M^* \left[
\, U - { e^{\pi/ka} \;\Gamma(A_1^*) \over 12 }\;M
\right] \ \ .\eqnum {A.6}
\end{equation}
In the above, $\;M,U \equiv M,U(A_1,4,2ikr)$, and $A_1$ was
defined in Eq. (A.3).  We now explain in some detail how
$c_{\,0}(k)$ and $c_{\,1}(k)$ are evaluated.

\subsection{Evaluation of $c_{\,0}(k)$.}
\label{subsec:A.1}
Using the identity:
\begin{equation}
M(a,c,2ikr) = {\Gamma(c) \over \Gamma(c-a) } e^{i\pi a} U(a,c,2ikr) +
 {\Gamma(c) \over \Gamma(a) } e^{i\pi(a-c)} e^{2ikr} U(c-a,c,-2ikr)  \ ,
\eqnum{A.6}
\end{equation}
and making a change of variable: $r=\rho/2k$, we rewrite Eq.
(A.4) as (below $\eta\equiv ka$):
\begin{eqnarray}
\nonumber
&& c_{\,0}(k) = \;\lambda\;
{ - i \; e^{-\pi/\eta} \over 2\,\eta \left| \; \Gamma(A_0)\right|^2 }
\;J_0 \ \ ,\\
\eqnum {A.7}
&& J_0 \equiv \int_{0}^{\infty} dr \; r \, \ln\left({r\mu \over 2k}
\right)
\left[ e^{-ir} U(A_0,2,i\,r)^2 \;\Gamma(A_0)^2 - {\rm {c.c.}} \right]\ \ ;
\end{eqnarray}
($\rm {c.c.} \equiv $ complex conjugate). If we now also
rewrite $U^2\,\Gamma(A_0)^2$ as
\begin{displaymath}
 U(A_0,2,ir)^2\;\Gamma(A_0)^2 = \left[ U(A_0,2,ir)^2\;\Gamma(A_0)^2
+ {1\over r^2} \right] \ - \ {1\over r^2} \ \ ,
\end{displaymath}
then the square bracket can be rotated to $r=-i\rho$ ($\rho
>0$) while its complex conjugate is rotated to $r=+i\rho$.
Eq. (A.7) becomes:
\begin{eqnarray}
\nonumber
J_0 =
&& 2i \int_{0}^{\infty} dr \; r \, \ln {r\mu \over 2k} \;
\;{\sin r \over r^2}\\
\eqnum {A.8}
&& \ - \int_{0}^{\infty} dr \; r \, e^{-r} \, \ln {r\mu \over 2k}\,
\left[ U(A_0,2,r)^2 \;\Gamma(A_0)^2 - {\rm {c.c.}} \right] \\
\nonumber
&& \ +{i\pi \over 2} \int_{0}^{\infty} dr \; r \, e^{-r} \,
\left[ U(A_0,2,r)^2 \;\Gamma(A_0)^2 + U(A_0^*,2,r)^2 \;\Gamma(A_0^*)^2
- {2\over r^2} \right] \ .
\end{eqnarray}
Note that each integral is convergent. If we define
\begin{equation}
 K_0(\epsilon) \equiv \int_{0}^{\infty} dr\; r^{1+\epsilon}\;e^{-r}\;
U(A_0,2,r)^2\;\Gamma(A_0)^2\ \ ,   \eqnum {A.9}
\end{equation}
(convergent for $\epsilon >0$), we have:
\begin{eqnarray}
\nonumber
J_0 = && i\pi \left( \ln{\mu \over 2k}  - \gamma_{E} \right)
\ -\ {\partial \over \partial \epsilon} \bigg[ K_0(\epsilon) - K_0(\epsilon)^*
\bigg]_{\epsilon=0^+} \\
\eqnum {A.10}
&& - \ln{\mu \over 2k}\;
  \bigg[ K_0(\epsilon) - K_0(\epsilon)^*\bigg]_{\epsilon=0^+}
+ {i\pi\over 2} \bigg[ K_0(\epsilon) + K_0(\epsilon)^* - 2\,\Gamma(\epsilon)
\bigg]_{\epsilon=0^+} \ .
\end{eqnarray}
To evaluate $K_0(\epsilon)$, we replace one of the $U$'s by
its integral representation
\begin{equation}
 e^{-r} \; U(a,c,r)\,\Gamma(a) = \int_0^{1} dt\; e^{-r/t}\,t^{-c}\,
(1-t)^{a-1} \ \ ,
\eqnum {A.11}
\end{equation}
and use the result
\begin{eqnarray}
\nonumber
&& \int_0^{\infty} dr\; r^{b-1}\,e^{-sr} U(a,c,r) = { \Gamma(b-c+1)
\over \Gamma(a) }\ s^{-b} \; \sum_{n=0}^{\infty}
{ \Gamma(a+n) \Gamma(b+n) \ \ \ (1-s^{-1})^n,
\over \Gamma(a+b-c+1+n) \Gamma(1+n)}  \ ,\\
\eqnum {A.12}
&& \hskip9truecm {\rm Re}(s) >1/2 \ .
\end{eqnarray}
We find:
\begin{equation}
 K_0(\epsilon) = \Gamma(1+\epsilon)^2 \; \sum_{n=0}^{\infty}
{ \Gamma(2+n+\epsilon) \, \Gamma(A_0+n)^2 \over \Gamma(1+n)
\,\Gamma(A_0+n+1+\epsilon)^2 } \ \ ,
\eqnum {A.13}
\end{equation}
which we rewrite for convenience as
\begin{equation}
 K_0(\epsilon) = \Gamma(1+\epsilon)^2 \left\{
\zeta(1+\epsilon)
+ \sum_{n=0}^{\infty} \left[
{ \Gamma(2+n+\epsilon) \, \Gamma(A_0+n)^2 \over \Gamma(1+n)
\,\Gamma(A_0+n+1+\epsilon) } - { 1 \over (n+1)^{1+\epsilon}}\right] \right\}
\eqnum {A.14}
\end{equation}
We would like to point out that the remaining sum is
convergent for $\epsilon=0$, and its derivative with respect
to $\epsilon$ also converges for $\epsilon=0$.  We then
obtain the following partial results:
\begin{eqnarray}
\nonumber
&&
{\partial \over \partial \epsilon}
\left[K_0(\epsilon) - K_0(\epsilon)^* \right]_{\epsilon=0} =
\sum_{n=0}^{\infty} \left[ {1+n \over (A_0+n)^2}
 \Big( \psi(n+2) -2\gamma_E -2\psi(A_0+n+1) \Big) - {\rm c.c.} \right],
\\
&&
\left[K_0(\epsilon) - K_0(\epsilon)^* \right]_{\epsilon=0} =
\sum_{n=0}^{\infty} \left[ {1+n \over (A_0+n)^2} - {1+n \over (A_0^*+n)^2}
\right] \ \ , \cr
\eqnum {A.15}
&&
\left[K_0(\epsilon) + K_0(\epsilon)^* -2\,\Gamma(\epsilon) \right]_
{\epsilon=0} =
\sum_{n=0}^{\infty} \left[ {1+n \over (A_0+n)^2} + {1+n \over (A_0^*+n)^2}
- {2\over n+1} \right].
\end{eqnarray}
Putting everything together, we find:
\begin{eqnarray}
 \nonumber
J_0 = &&
 - {i\pi \over 2} \Big( 2\,\gamma_E + \psi(A_0) + \psi(A_0^*)
+(A_0-1)\,\psi^{\,'}(A_0)
+ (A_0^*-1)\,\psi^{\,'}(A_0^*) \Big) \\
\eqnum {A.16}
&& +  \left( \ln{\mu \over 2k} - 2\, \gamma_E  \right)
\Big( i\pi + \psi(A_0) - \psi(A_0^*) +(A_0-1)\,\psi^{\,'}(A_0)
	- (A_0^*-1)\,\psi^{\,'}(A_0^*) \Big) \\
\nonumber
&& + \sum_{n=0}^{\infty} \left[ {1+n \over (A_0+n)^2}
 \Big( 2\psi(A_0+n+1) - \psi(n+2) \Big) - {\rm c.c.} \right] \ \ .
\end{eqnarray}
For small velocities ($i.e.$ for $\eta=ka
\;\lower2pt\hbox{$\lesssim$}\; 0.1$), we can use the asymptotic
behavior of the $\psi$ function and its derivatives to rewrite Eq.
(A.16) as
\begin{equation}
 J_0 = 2i\pi \;\left(
\ln{\mu a \over 2} -1 -2\,\gamma_E + {\eta^2 \over 12} + {\eta^4 \over 40}
+ \dots \right)  \ \ ,
\eqnum {A.17}
\end{equation}
and our result for $c_{\,0}(k)$ is:
\begin{equation}
 c_{\,0}(k) =  {\lambda \over 2} \,\left[ \ln {\mu\,a\over 2} - 1 - 2\,\Euler
+ {\eta^2 \over 12} + {\eta^4 \over 40} + \dots \right] \ \ .
\eqnum {A.18}
\end{equation}
For $ \eta > 0.1$, the sum in Eq. (A.16) is evaluated
numerically.  See Fig. 1 for a plot of $c_{\,0}(k)$ as a
function of $\eta$, when $\mu=2/a$.

\subsection{Evaluation of $c_{\,1}(k)$}
\label{subsec:A.2}

For the case $\ell=1$, we need to evaluate
\begin{eqnarray}
 \nonumber
&& c_{\,1}(k) =\;\lambda \;
{ - i \; e^{-\pi/\eta} \over 2\,\eta \left| \;
\Gamma(A_1)\right|^2 } \;J_1 \ \ ,
\\
\eqnum {A.19}
&& J_1 \equiv \int_{0}^{\infty} dr \; r^3 \, \ln\left({r\mu \over 2k}\right)
\left[ e^{-ir} U(A_1,4,ir)^2 \;\Gamma(A_1)^2 - {\rm {c.c.}} \right]\ \ .
\end{eqnarray}
$J_1$ is evaluated in the same maner as $J_0$.  We first
rewrite $U^2\Gamma(A_1)^2$ as
\begin{eqnarray}
 \nonumber
&& U(A_1,4,i\,r)^2\;\Gamma(A_1)^2 = \left[ U(A_1,4,i\,r)^2\;\Gamma(A_1)^2
-{\cal{D}} (ir) \right] + {\cal {D}} (ir) \ \ ,\\
\eqnum {A.20}
&& {\cal{D}}(z) \equiv {4 \over z^6} + { 12-4\,A_1^2 \over z^5}
+ { 21 - 16\,A_1 + 3\,A_1^2 \over z^4} \ \ .
\end{eqnarray}
Then the square bracket is rotated to $r=-i\rho$, and its
complex conjugate to $r=i\rho$. Eq. (A.19) becomes:
\begin{eqnarray}
\nonumber
J_1 = && \int_{0}^{\infty} dr \; r \, \ln {r\mu \over 2k} \;
\left[ e^{-ir} \,{\cal {D}} (ir) - {\rm c.c.} \right] \\
\eqnum {A.21}
&& \ + \int_{0}^{\infty} dr \; r \, e^{-r} \, \ln {r\mu \over 2k} \;
\left[ U(A_1,4,r)^2 \;\Gamma(A_1)^2 - {\cal{D}}(r) - {\rm {c.c.}} \right] \\
\nonumber
&& \ - {i\pi \over 2} \int_{0}^{\infty} dr \; r \, e^{-r} \,
\left[ U(A_1,4,r)^2 \;\Gamma(A_1)^2 - {\cal{D}}(r) + {\rm c.c.}
\right] \ \ ,
\end{eqnarray}
where each integral is convergent.  The first integral is
easily done and gives:
\begin{equation}
 {\cal{R}} \equiv i \left( \ln { \mu \over 2k } - \gamma_E \right)
\; \left( { 3\pi \over \eta^2 } + {8 \over \eta} + \pi \right)
+ i\,\left( {8 \over \eta} + \pi \right) \ \ .\eqnum {A.22}
\end{equation}
To compute the remaining integrals, we define
\begin{eqnarray}
\nonumber
&& K_1(\epsilon) \equiv \int_{0}^{\infty} dr\; r^{3+\epsilon}\;e^{-r}\;
U(A_1,4,r)^2\;\Gamma(A_1)^2\ \ ,\\
\eqnum {A.23}
&& L_1(\epsilon) \equiv \int_{0}^{\infty} dr\; r^{3+\epsilon}\;e^{-r}\;
{\cal{D}}(r) \ \ ,
\end{eqnarray}
which are convergent for $\epsilon >2$. Eq. (16) is then
rewritten as:
\begin{eqnarray}
\nonumber
J_1 = {\cal{R}} + \lim_{\epsilon \to 0} &&
\Bigg[ \Big( \; {\partial \over \partial \epsilon} + \ln{\mu \over 2k}
\;\Big)
\; \Big(\; K_1(\epsilon) - K_1(\epsilon)^*
- L_1(\epsilon) + L_1(\epsilon)^* \;\Big) \\
\eqnum {A.24}
&& - {i\pi\over 2}  \Big( \;K_1(\epsilon) + K_1(\epsilon)^* - L_1(\epsilon)
- L_1(\epsilon)^* \;\Big) \Bigg] \ \ .
\end{eqnarray}
Note that the result for $\epsilon=0$ is obtained by
analytical continuation. The results for $K_1$ and $L_1$ are
easily obtained if one uses Eqs. (A.11) and (A.12):
\begin{eqnarray}
 \nonumber
&& L_1(\epsilon) = 4\,\Gamma(-2+\epsilon) +(12-4\,A_1)\,\Gamma(-1+\epsilon)
+(21-16\,A_1 +3\,A_1^2)\,\Gamma(\epsilon) \ \ ,\\
\eqnum {A.25}
&& K_1(\epsilon) = \Gamma(1+\epsilon)^2 \; \sum_{n=0}^{\infty}
{ \Gamma(4+n+\epsilon) \, \Gamma(A_1+n)^2 \over \Gamma(1+n)
\,\Gamma(A_1+n+1+\epsilon)^2 } \ \ .
\end{eqnarray}
To extract the pole part of $K_1(\epsilon)$, we rewrite the last equation as:
\begin{equation}
 K_1(\epsilon) = \Gamma(1+\epsilon)^2\, \sum_{n=0}^{\infty} u_n(\epsilon)
+ \Gamma(1+\epsilon)^2  \sum_{n=0}^{\infty}
\left[ { \Gamma(4+n+\epsilon) \, \Gamma(A_1+n)^2 \over \Gamma(1+n)
\,\Gamma(A_1+n+1+\epsilon)^2 } - u_n(\epsilon) \right] \;,
\eqnum {A.26}
\end{equation}
where
\begin{eqnarray}
\nonumber
u_n(\epsilon) \equiv && {1 \over (n+1)^{-1+\epsilon} }
+{ 5-2\,A_1 +\epsilon (7/2-2\,A_1) \over (n+1)^{\epsilon} } \\
\eqnum {A.27}
&& \ \ \ \
+{ 11-12A_1 +3A_1^2 + \epsilon (175/12-18A_1+5A_1^2) \over (n+1)^{1+\epsilon}}
\ \ .
\end{eqnarray}
The first sum which contains the pole part gives
\begin{eqnarray}
 \nonumber
\sum_{n=0}^{\infty} u_n(\epsilon) =
&& \zeta(-1+\epsilon)
 +\zeta(\epsilon) \;\left[ 5-2\,A_1 +\epsilon \left({7\over 2}-2\,A_1\right)
\right] \\
\eqnum {A.28}
&& \ \ \ \
+\zeta(1+\epsilon) \left[ 11-12\,A_1 +3\,A_1^2
+ \epsilon \left({175 \over 12} - 18\, A_1 + 5\,A_1^2\right) \right] \ .
\end{eqnarray}
In the above equation, $\zeta$ is the Riemann's zeta
function.  Now the second sum in Eq. (A.26) converges for
$\epsilon=0$ and so does its derivative with respect to
$\epsilon$. We obtain the following partial results:
\begin{eqnarray}
 \nonumber
&& \sum_{n=0}^{\infty}
\left[ { \Gamma(4+n+\epsilon) \, \Gamma(A_1+n)^2 \over \Gamma(1+n)
\,\Gamma(A_1+n+1+\epsilon)^2 } - u_n(\epsilon) \right]\ \Bigg|
_{\epsilon \to 0} \\
\eqnum {A.29}
&& \ \ \ \ \ \ \ \ = \ i\,\left( {1 \over \eta^3} + {1\over \eta} \right)\;
\psi^{\,'}(A_1) + \left( 1 + { 3\over \eta^2} \right) \;
\bigg( \psi(A_1) + \gamma_E \bigg) \ \ ,
\end{eqnarray}
\begin{eqnarray}
\nonumber
&& {\partial \over \partial \epsilon} \sum_{n=0}^{\infty}
\left[ { \Gamma(4+n+\epsilon) \, \Gamma(A_1+n)^2 \over \Gamma(1+n)
\,\Gamma(A_1+n+1+\epsilon)^2 } - u_n(\epsilon) - {\rm c.c.} \right]\ \Bigg|
_{\epsilon \to 0} = \\
\nonumber
&& \ \ \
\left(2+{5\over \eta^2} \right) \; \bigg(\psi(A_1) - \psi(A_1^*) \bigg)
 - {9\,i\over \eta } \;\bigg( \psi(A_1) + \psi(A_1^*) \bigg)
- { 2\,i\over \eta } \bigg( \ln 2 \pi - {21\over 2} + 8\,\gamma_E \bigg) \\
\eqnum {A.30}
&& \ \ \
+ i\,\left( {1 \over \eta^3} + {1\over \eta} \right)\;
\sum_{n=0}^{\infty} \left[
 { \psi(n+4) -2\psi(A_1+n+1) \over (A_1+n)^2} + {\rm c.c.} \right] \\
\nonumber
&& \ \ \
- \left( 1 + { 3\over \eta^2} \right) \; \sum_{n=0}^{\infty} \left[
 { \psi(n+4) -2\psi(A_1+n+1) \over A_1+n } - {\rm c.c.} \right] \ \ ,
\end{eqnarray}
where we have used $A_1=2+i/\eta$. For small velocities, we
have:
\begin{equation}
 J_1 = 2\,i\,\pi \left[
\left( \ln {\mu a\over 2} - 2\,\gamma_E \right)\,
\left( 1+{3\over \eta^2} \right) + {4 \over \eta^2} - {1\over 12} +
{11 \over 120} \,\eta^2 - {1\over \eta\,\pi}
\right] + {\cal{O}}(\eta^4) \ ,
\eqnum {A.31}
\end{equation}
and our result for $c_{\,1}(k)$, $\eta \;\lower2pt\hbox{$\lesssim$}\;
0.1$, is:
\begin{eqnarray}
\nonumber
c_{\,1}(k) = \lambda \Bigg[\
\left( \ln {\mu a\over 2} - 2\,\gamma_E \right)\,
&& \left( -\,{3\over 2} + \eta^2 - \eta^4 \right)
- 2 + {49\over 24}\,\eta^2  - {167 \over 80} \,\eta^4 \\
 \eqnum {A.32}
&& + {1\over 2\,\pi} \left( \eta - \eta^3 + \eta^5 \right)
+ {\cal{O}}(\eta^6) \ \Bigg] \ \ .
\end{eqnarray}
See Fig. 2 for a plot of $c_{\,1}(k)$ as a function of
$\eta$, when $\mu=2/a$.

Before we close this section, we would like to add a few words about
the correctness of the results presented here.  We have checked that
the terms which are proportional to $\;\ln\,\mu\;$ in Eqs. (A.19,
A.20) are the same as those obtained by solving Eq. (A.1) directly,
with the obvious replacement $\;\ln\,\mu r \rightarrow \ln\,\mu\;$;
they can be deduced from Eq. (A.2), with $r=0$, and
$\;a^{-1}\rightarrow a^{-1} + a^{-1}\lambda\ln\,\mu\;$.  These terms
are needed to show that the physical results (see for instance Eq.
(3.9)) are, to the order we are working, $\mu$-independent (this is of
course a consequence of the $\mu$-independence of the Hamiltonian $H$
in Eq. (3.2)).  We have also checked that Eqs. (A.18, A.32) reproduce
the correct results in the limit $ k \rightarrow 0$. The latter are
obtained as follows: one first regulates the integrals in Eqs. (A.7,
A.19) (e.g., by adding a term $e^{-r\epsilon}$ in the integrands),
then one uses
\begin{displaymath}
\lim_{a \to \infty} \;\Gamma(a-c+1)\;U(a,\;c,\;z/a) =
2\;z^{1/2-c/2}\;K_{c-1} \left(2\,z^{1/2}\right) \ \ ,
\end{displaymath}
where $K_{c-1}$ is the Bessel function of the second kind.  The
integrals one obtains involve a product of Bessel functions, and are
easily done. This provides a crucial test of the correctness of the
calculation.



\newpage
\begin{figure}[b]
\begin{center}
    \leavevmode
    \epsfverbosetrue
    \epsfxsize=4truein
    \caption{}
\end{center}
\end{figure}

\begin{center}
$c_0$ versus $\eta\equiv ka$.
\end{center}

\newpage
\begin{figure}[b]
\begin{center}
    \leavevmode
    \epsfverbosetrue
    \epsfxsize=4truein
\caption{}
\end{center}
\end{figure}

\begin{center}
 $c_1$ versus $\eta\equiv ka$.
\end{center}

\newpage
\begin{figure}[b]
\begin{center}
    \leavevmode
    \epsfverbosetrue
    \epsfxsize=4truein
\caption{}
\end{center}
\end{figure}

\vbox{$R_c^V$ v.s. $\sqrt{s}$, with $m_c=1.57\ GeV$,
$\Lambda_{QCD}=0.2\ GeV$, $\langle \alpha_{\hskip-1.3pt
s}\,G^2\rangle =0.042\ GeV^4$.  The solid and dotted lines
were obtained with the choice $\mu=2\,m_c$; the dashed-line
was obtained by choosing $\mu$ such that the radiative
corrections vanish ($i.e.$ such that $c_0=0$).  This shows
that $R_c^V$ is not very sensitive to the choice of $\mu$.}

\newpage
\begin{figure}[b]
\begin{center}
    \leavevmode
    \epsfverbosetrue
    \epsfxsize=4truein
\caption{}
\end{center}
\end{figure}

\vbox{Theoretical predictions for $R_c^V$ together with
experiment results (Ref. [15], Particle Data Book). The
dotted-line represents the parton model prediction, and is
obtained with $m_c=1.57\ GeV$, $\mu=2\,m_c$,
$\Lambda_{QCD}=0.2\ GeV$, and $\langle \alpha_{\hskip-1.3pt
s}\,G^2\rangle =0.042\ GeV^4$.  The solid line represents
the ``exact'' value of $R_c^V$ and the gray area surrounding it
represents the uncertainties which are obtained by varying
$m_c$, $\Lambda_{QCD}$ and $\langle \alpha_{\hskip-1.3pt
s}\,G^2\rangle $. The dashed-line gives $R_c$ when radiative
and nonperturbative corrections are neglected.  The
experimental data is represented by the diamonds; the
horizontal error is less than 4\% while the vertical errors
are large ($\sim 5-35\%$).}

\newpage
\begin{figure}[b]
\begin{center}
    \leavevmode
    \epsfverbosetrue
    \epsfxsize=4truein
\caption{}
\end{center}
\end{figure}

\centerline{$R_b^V$ v.s. $\sqrt{s}$. Same conventions as in Fig. 4.}

\newpage
\begin{figure}[b]
\begin{center}
    \leavevmode
    \epsfverbosetrue
    \epsfxsize=4truein
\caption{}
\end{center}
\end{figure}

\vbox{Prediction for $R_t^V$ v.s. $\sqrt{s}$, with
$\mu=\,M_Z$, $\alpha_{\hskip-1.pt s}=0.119$, and $\langle
\alpha_{\hskip-1.pt s}\,G^2\rangle =0.042\ GeV^4$.  $R_t^{V
(exact)}$ is plotted for $m_t=165\ GeV$ (solid line),
$m_t=180\ GeV$ and $m_t=150\ GeV$ (dashed lines).}

\newpage
\begin{figure}[b]
\begin{center}
    \leavevmode
    \epsfverbosetrue
    \epsfxsize=4truein
\caption{}
\end{center}
\end{figure}

Prediction for $R_t^A$ v.s. $\sqrt{s}$, with $m_t=165\ GeV$,
and $\langle \alpha_{\hskip-1.pt s}\,G^2\rangle =0.042\
GeV^4$.  We plot $R_t^{V (exact)}$ using two different
criterias for the choice of $\mu$: (a) choose $\mu=\,M_Z$
and $\alpha_{\hskip-1.pt s}=0.119$ (solid line); (b) choose
$\mu$ such that the radiative corrections vanish, $i.e.$
such that $c_1=0$ (dashed line).

\end{document}